\documentclass[preprint,aps,prd,showpacs,nofootinbib,superscriptaddress,floatfix,dvipdfmx]{revtex4-1}

\usepackage{amsmath}
\usepackage{amssymb}
\usepackage{graphicx}
\usepackage{ascmac}
\usepackage{url}

\usepackage{hyperref}
\hypersetup{%
 setpagesize=false,%
 bookmarks=true,%
 bookmarksdepth=tocdepth,%
 bookmarksnumbered=true,%
 colorlinks=true,%
 allcolors=blue,
 pdftitle={},%
 pdfsubject={},%
 pdfauthor={},%
 pdfkeywords={}}
 
\def\VEV#1{
\langle #1\rangle
}
\begin{document}
\title{Does finetuning make $D$-term contributions smaller in  natural grand unified theories with spontaneous supersymmetry breaking?}

\author{Nobuhiro Maekawa}
\email[]{maekawa@eken.phys.nagoya-u.ac.jp}
\affiliation{Department of Physics,
Nagoya University, Nagoya 464-8602, Japan}
\affiliation{Kobayashi-Maskawa Institute for the Origin of Particles and the
Universe, Nagoya University, Nagoya 464-8602, Japan}

\author{Taiju Tanii}
\email[]{tanii.t@eken.phys.nagoya-u.ac.jp}
\affiliation{Department of Physics,
Nagoya University, Nagoya 464-8602, Japan}

\date{\today}

\begin{abstract}
    \noindent
In this paper, we explore the natural grand unified theory (GUT) with spontaneous supersymmetry (SUSY) breaking,  focusing on the contribution to sfermion and Higgs masses. Natural GUTs solve various problems in SUSY GUTs with the natural assumption that all terms allowed by $SO(10)\times U(1)_A$ symmetry are introduced with $O(1)$ coefficients. 
It is also possible to introduce spontaneous SUSY breaking in the natural GUT. This scenario predicts $D$-term contribution to sfermion and Higgs masses dominates the $F$-term contribution, that potentially leads to the SUSY flavor problem. Fortunately, it also predicts high-scale SUSY, which avoids the SUSY flavor problem but leads to the instability of the 
electroweak scale. The $D$-term domination results in the superheavy Higgsino unless the $D$-term contribution becomes comparable with the $F$-term contribution.
We discuss whether it is possible to suppress the $D$-term contribution while maintaining the $F$-term contribution through the tuning of $O(1)$ coefficients in the natural GUT with spontaneous SUSY breaking. Our results indicate that it is impossible. 
This also suggests that the mass spectrum of the sfermions can be predicted by the $D$-term contributions, and that the Higgsino is not a candidate for dark matter in the scenario.

\end{abstract}

\maketitle
\section{Introduction}
The grand unified theory (GUT) \cite{Georgi:1974sy} is one of the most promising candidates for a model beyond the standard model (SM).
In the GUT, the three gauge interactions in the SM, $SU(3)_C\times SU(2)_L\times U(1)_Y$, are unified into one gauge interaction, for example, $SU(5)$, $SO(10)$, and $E_6$. It also realizes the unification of quarks and leptons, for example, one generation quarks and leptons can be unified into $\bf 10$ and $\bf\bar 5$ of $SU(5)$, or $\bf 16$ of $SO(10)$.
It is surprising that we have already known the experimental supports for both unifications.
Regarding the unification of matter, the various mass hierarchies observed in experiments can be explained only by one assumption that  $\bf 10$ fields of $SU(5)$ induce larger hierarchy in Yukawa couplings than $\bf\bar 5$ fields of $SU(5)$. 
As for the unification of gauge interactions, it has been shown that, by introducing supersymmetry (SUSY) \cite{SUSY,SUSYGUT}, the three gauge coupling constants of the Standard Model coincide at the grand unified scale $\Lambda_G\sim 2\times 10^{16}$ GeV \cite{threegaugeunification}.

However, there are two problems in simple SUSY GUTs \cite{SUSYGUT}.
The first problem is that, due to the unification of Standard Model particles, it leads to the unrealistic result that their masses are equal.
The second is that, when unifying the SM Higgs into the $\bf 5$ representation of $SU(5)$, the mass of its partner (called colored Higgs or triplet Higgs) must be significantly larger than that of the SM Higgs to ensure stable protons.
The latter is known as the doublet-triplet (DT) splitting problem \cite{Randall:1995sh}.

The natural GUTs \cite{NaturalSO10,NaturalE6} have a natural assumption that all terms allowed by the symmetry are introduced with $O(1)$ coefficients, thereby solving the above problems in SUSY GUT.
Furthermore, the natural GUT can naturally introduce the mechanism for spontaneous SUSY breaking \cite{Fayet:1974jb,Kim:2008kw, Maekawa:2017cva,Maekawa:2019vzk} and predicts high scale SUSY at around $1000$ TeV. This has the advantage of suppressing flavor changing neutral current (FCNC) processes and CP-violating processes, mediated by SUSY particles, but also has the disadvantage that the electroweak scale is unstable \cite{Arkani-Hamed:2012fhg}.

As discussed in detail in this paper, there are two types of contributions to the sfermion and Higgs masses, and the fact that the $D$-term contribution becomes one order of magnitude larger than the $F$-term contribution. 
Specifically, a large Higgsino mass $\mu$ is required to cancel the large negative $D$-term contribution to the squared mass of Higgs, which predicts that the Higgsino is a superheavy particle and therefore, it cannot be the dark matter. If the $D$-term contribution is canceled by the $F$-term contribution, the Higgsino mass can be around the weak scale and therefore the Higgsino can be the dark matter. 
To realize the cancellation, the $D$-term contribution must be comparable with the $F$-term contribution.

In this paper, we discuss whether it is possible to overturn the dominance of the $D$-term contribution. Specifically, we explore the possibility of suppressing only the $D$-term contribution while maintaining the $F$-term contribution by tuning several $O(1)$ coefficients. Unfortunately, our conclusion is that the $D$-term dominance is quite a general result in this scenario. And therefore, the sfermion mass spectrum is fixed by the $D$-terms, i.e., the quantum numbers of quarks, leptons, and (GUT) Higgses, and the Higgsino cannot be the dark matter.

After this introduction, we review the natural GUT and the mechanism for spontaneous SUSY breaking in section II. In section III, we use a simple model to concretely demonstrate the relationship between $F$-term contribution and $D$-term contribution. We conclude that the $D$-term contribution cannot be comparable with the $F$-term contribution by the cancellation between parameters. 
In a more general model, the conclusion does not change. 
Section IV is devoted to discussions and summary.

\section{Natural GUT and SUSY breaking}

In this section,  
we review the natural $SO(10)$ GUT \cite{NaturalSO10} and the spontaneous SUSY breaking \cite{Maekawa:2019vzk}.

The anomalous $U(1)_A$ gauge symmetry plays an essential role to restrict the interactions in the superpotential.
One of the most important features in the natural GUT is that all terms allowed by the symmetry are introduced with $O(1)$ coefficients.  
The typical quantum numbers for the field content under the symmetry $SO(10)\times U(1)_A\times Z_2$ are summarized in Table \ref{SO10}. The typical model can solve the DT splitting problem and obtain realistic quark and lepton masses and mixings.
\begin{table}[t]
  \begin{center}
    \begin{tabular}{|c||c|c|c|}
      \hline
      $SO(10)$ & negatively charged fields & positively charged fields & matter fields \\ \hline \hline
      {\bf 45}  & $A(a=-1,-)$ & $A'(a'=3,-)$ & \\ \hline
      {\bf 16} & $C(c=-4,+)$ & $C'(c'=3,-)$  & $\Psi_i(\psi_1=\frac{9}{2}, \psi_2=\frac{7}{2}, \psi_3=\frac{3}{2}, +)$ \\ \hline
      ${\bf \overline{16}}$ & $\bar C(\bar c=-1,+)$ & $\bar C'(\bar c'=7,-)$ & \\ \hline
      {\bf 10} & $H(h=-3,+)$  & $H'(h'=4,-)$ & $T(t=\frac{5}{2},+)$ \\ \hline
      1 &  $\left.\begin{array}{c}\Theta(\theta=-1,+), \\ Z(z=-2,-), \bar Z(\bar z=-2,-) \end{array}\right.$ & $S(s=5,+)$ &
      \\ \hline
    \end{tabular}
    \caption{Field content of natural $SO(10)$ GUT with $U(1)_A$ charges. $\pm$ labels the $Z_2$ parity. }
    \label{SO10}
  \end{center}
\end{table}
Here, $Z_2$ symmetry is introduced to realize the DT splitting as discussed in Ref.\cite{NaturalSO10}.

In this paper, we use large characters for superfields or operators and small characters for their $U(1)_A$ charges.
The $U(1)_A$ gauge symmetry \cite{U(1)} has gauge anomalies, but they can be cancelled by the Green-Schwarz mechanism \cite{Green-Schwartz}. Furthermore, the Fayet-Iliopoulos (FI) term \cite{Fayet:1974jb} $\xi^2\int d^2\theta V_A$ is assumed, with a vector multiplet $V_A$ of $U(1)_A$ and a small constant parameter $\xi\sim 0.22\Lambda$, where $\Lambda$ is a cutoff.

\subsection{Anomalous $U(1)_A$ gauge symmetry}
We consider a simpler model in which we have only three matter fields
$\Psi_i$, and two negatively charged fields $H$ and $\Theta$ as shown in Table \ref{SO10} for simplicity. 
The superpotential invariant under $U(1)_A$ is given as
\begin{equation}
\label{SO10Y}
W=y_{ij}\sum_{i,j=1,2,3}\left(\frac{\Theta}{\Lambda}\right)^{\psi_i+\psi_j+h}\Psi_i\Psi_jH,
\end{equation}
where $y_{ij}$ 
are the $O(1)$ coefficients. 
The cutoff scale is determined by gauge coupling unification (GCU) condition and becomes  the usual GUT scale $\Lambda_G\sim2\times10^{16}\rm{GeV}$ (Here, we do not explain in detail, please look at \cite{NaturalSO10}).
The GCU conditions also require $h\sim0$.
We assume that only $\Theta$ has a non-vanishing vacuum expectation value (VEV) here. It is determined as $\langle\Theta\rangle=\xi\equiv\lambda\Lambda$ by the $D$-flatness condition of $U(1)_A$.
Then the interaction terms in the above superpotential become the Yukawa interactions as
\begin{equation}
W=y_{ij}\lambda^{\psi_i+\psi_j+h}\Psi_i\Psi_jH.
\end{equation}
In this paper, we take $\lambda\sim 0.22$, which is approximately the Cabibbo angle. 
Unless otherwise noted, the O(1) coefficients like $y_{ij}$ are omitted and we take $\Lambda=1$ for simplicity in this paper.
The mechanism that realizes the hierarchical structure of Yukawa couplings by a field that obtains a VEV is called the Froggatt-Nielsen mechanism \cite{Froggatt-Nielsen} in the literature. 

Furthermore, the holomorphic feature of the superpotential forbids some higher dimensional terms. 
This mechanism is called the SUSY zero mechanism, and it has the property that only interactions with non-negative powers of $\lambda$ are allowed.

\subsection{Vacuun expectation values}
We explain how the VEVs of Higgs fields are determined by the $U(1)_A$ symmetry. For simplicity, we introduce only gauge singlet fields $Z_i^\pm(i=1,2,\dots n_\pm)$ with $U(1)_A$ charges $z_i^\pm(z_i^+>0 $ \rm{and} $ z_i^-<0)$. The VEVs of Higgs fields can be determined by $n_++n_-$ $F$-flatness conditions and one $D$-flatness condition,
\begin{equation}
F^\pm_i=-\left(\frac{\delta W}{\delta Z^\pm_i}\right)^*=0, \quad D_A=g_A\left(\sum_{i,\pm} z^\pm_i\left|Z^\pm_i\right|^2+\xi^2\right)=0,
\end{equation}
where, $g_A$ is the gauge coupling of $U(1)_A$. Due to the gauge invariance of the superpotential, the $F$-flatness conditions are no longer independent, and one condition is reduced. With generic interactions, we have always SUSY vacuum with $O(1)$ VEVs. 
However, in case $n_+< n_-$, completely different vacuum, where all positively charged fields have vanishing VEVs and only negatively charged fields have small VEVs, appear. 
Since each term in the superpotential contains at least one positively charged field, giving the positively charged fields vanishing VEVs automatically satisfies $F$-flatness conditions of negatively charged fields.
Also, the VEVs of negatively charged fields can be determined by the $n_+$ $F$-flatness conditions of positively charged fields and a $D$-flatness condition.
If $n_++1<n_-$, i.e., the number of constraints is smaller than that of the variables, it is impossible to determine the VEVs of all negatively charged fields, leading to the appearance of massless fields. When $n_++1=n_-$, the number of constraints are balanced with that of the variables, and all negatively charged fields are determined as
\cite{NaturalSO10}
\begin{equation}
\label{VEV}
\VEV{Z}\sim \left\{\begin{array}{l}
\lambda^{-z}\quad (z\leq 0)\cr
0\qquad (z>0)
                       \end{array}\right..
\end{equation}
On the other hand, in the case of $n_++1>n_-$, the number of constraints becomes larger than that of the variables, we have no SUSY vacuum with vanishing VEVs of positively charged fields, and therefore, 
SUSY is spontaneously broken \cite{Kim:2008kw}. The potential energy of the SUSY breaking vacuum becomes $V_{SB}\sim |F_{S_{max}}|^2\sim \lambda^{2s_{max}}\Lambda^4$, where the field $S_{max}$ has the maximal $U(1)_A$ charge. Since the $V_{SB}$ is much smaller than the potential barrier between this SUSY breaking vacuum and the SUSY vauum with $O(1)$ VEVs when $s_{max}$ is sufficiently large, the meta-stable vacuum can have longer life-time than the age of the universe.

Note that terms that include more than one positively charge field can be ignored in determining the vacuum because of vanishing VEVs of positively charge fields. 
Therefore, the superpotential $W_{Z^+}$, which includes only one positively charged field $Z^+$, is sufficient to be considered when the VEVs are determined. This is important not only in determining the VEVs but also in considering SUSY breaking. 
Assigning a $U(1)_R$ charge of 2 to positively charged fields and 0 to negatively charged fields ensures that the superpotential determining the VEV respects the $U(1)_R$ symmetry. Moreover, the non-vanishing VEVs of negatively charged fields and $F$ of positively charge fields do not break the $U(1)_R$. Therefore, the SUSY breaking parameters which break $U(1)_R$, the gaugino masses and $A$ parameters, become much smaller than the SUSY breaking parameters which do not break $U(1)_R$, the sfermion masses.

\subsection{Spontaneous SUSY breaking}
In the natural GUT, VEVs of all negatively charged fields except a pair of the Higgs doublet fields are determined by the $F$-flatness conditions of positively charged fields. If we add one positively charged field or omit one negatively charged field, SUSY is spontaneously broken. 
The easiest way to build such a model is to omit one negatively charged field $\bar Z$ from the natural GUT in Table \ref{SO10} \cite{Maekawa:2019vzk}. 
Thus, one of the $F$-flatness conditions of $C'$ and $\bar C'$, which are obtained by the 
superpotentials:
\begin{eqnarray}
W_{\bar{C}^{\prime}}&=&\bar{C}^{\prime}\left(\lambda^{\bar{c}^{\prime}+c+a} A+\lambda^{\bar{c}^{\prime}+c+z} Z\right) C,\cr
W_{{C}^{\prime}}&=&\bar{C}\left(\lambda^{{c}^{\prime}+{\bar c}+a} A+\lambda^{{c}^{\prime}+{\bar c}+z} Z\right) C',
\end{eqnarray}
cannot be satisfied, and therefore, SUSY is spontaneously broken. Here, the VEV of $A$ is fixed by $F_{A'}=0$ and therefore, choosing the VEV of $Z$ cannot satisfy both $F_{C'}=0$ and $F_{\bar C'}=0$.
When we take larger $\bar c'$, the SUSY breaking scale can be obtain as
\begin{equation}
m_{S U S Y}=\frac{F_{\bar{C}^{\prime}}}{\Lambda} \sim \lambda^{\bar{c}^{\prime}+\frac{1}{2} (c-\bar{c})} \Lambda,
\end{equation}
because $\VEV{A}\sim\lambda^{-a}$,  
$\VEV{Z}\sim\lambda^{-z}$, 
and $|\VEV{C}|=|\VEV{\bar C}|\sim \lambda^{-\frac{1}{2}(c+\bar c)}$, where $|\VEV{C}|=|\VEV{\bar C}|$ is required by $D$-flatness condition of $SO(10)$. 
\footnote{
As discussed in ref. \cite{Maekawa:2017cva}, the term $\bar C'AH^2C$ must be forbidden by some discrete symmetry because
$AH^2$ can play the same role as $\bar Z$ and therefore, SUSY vacuum appears. }

If SUGRA effects are not taken into account, 
the gaugino masses become $m_{1/2} \sim m_{\rm{SUSY }}^2/\Lambda$, which is much smaller than $m_{\rm{SUSY}}$, because for non-vanishing gaugino masses, small $U(1)_R$ breaking must be picked up. Fortunately, the SUGRA effect \cite{Maekawa:2017cva} gives larger, but still small, gaugino masses which are comparable to the gravitino masses as
\begin{equation}
m_{1/2}\sim m_{3/2}=\frac{F_{\bar{C}^{\prime}}}{M_{\rm{Pl}}}\sim m_{\rm{SUSY }}\frac{\Lambda}{M_{\rm{Pl}}} \sim 10^{-2} m_{\rm{SUSY }}, 
\end{equation}
where $M_{\rm{Pl}}\sim 2\times10^{18}$ GeV is a reduced Planck scale.

For the sfermion masses, this model has two contributions: $F$-term contributions through the higher dimensional term, such as $\int d^4 \theta\left|\bar{C}^{\prime}\right|^2 \Psi_i^{\dagger} \Psi_i$ and the $D$-term contribution from $\int d^4 \theta \Psi_i^{\dagger} e^{\psi_i g_A V_A} \Psi_i$. 
As discussed below in the simplest model, the $D$-term contribution is 10 times larger than the $F$-term contribution.

We will show the result with the simplest model discussed in \cite{Kim:2008kw}.
We introduce one positively charged field $S$ and one negatively charged field $\Theta$. 
The generic superpotential and scalar potential are given as
\begin{eqnarray}
\label{W}
    & W= f(S\Theta^s) , \\
& V=\left| F_S\right|^2+\left|F_{\Theta}\right|^2+\frac{1}{2} D_A^2,
\end{eqnarray}
where $f(x)$ is a function of $x$, $F_S^*=-f'(S\Theta^s)\Theta^s$, $F_\Theta^*=-sf'(S\Theta^s)S\Theta^{s-1}$,
and $D_A=g_A(s|S|^2-|\Theta|^2+\xi^2)$.
This model has SUSY vacua, where $f'(S\Theta^s)=0$ and $D_A=0$ are satisfied. But if $s\gg 1$, the meta-stable vacuum, where $\VEV{S}\sim 0$ and $\VEV{\Theta}\sim \xi\Lambda$, appears. 
Basically, the meta-stable vacuum is understood by solving the stationary conditions $\frac{\partial V}{\partial S}=0$ and
$\frac{\partial V}{\partial \Theta}=0$.
$\frac{\partial V}{\partial S}=0$ determines the VEV of $S$ and
$\frac{\partial V}{\partial \Theta}=0$
determines the VEV of $\Theta$. The VEV of $\Theta$ is approximately determined by $D_A=0$, but correct value is determining by fixing $D$-term
by the stationary condition as  
\begin{equation}
\label{Dfix}
\frac{\partial V}{\partial \Theta}=  F_S \frac{\partial F_S^*}{\partial \Theta}+F_\Theta \frac{\partial F_\Theta^*}{\partial \Theta}+D_A \frac{\partial D_A}{\partial \Theta}=0.
\end{equation}
We can ignore the second term, because $F_\Theta$ include $S$, which has small VEV. 
Thus, we obtain the relation $g_AD_A=s\left| F_S\right|^2|\Theta|^{-2}$ because 
$\frac{\partial F_S^*}{\partial \Theta}\sim sF_S^*\Theta^{-1}$ and $\frac{\partial D_A}{\partial \Theta}\sim -g_A\Theta^*$.
Therefore, we conclude that $D$-term contribution is 10 times larger than the $F$-term contribution since
$s\sim O(10)$ and $|\Theta|^{-2}\sim O(10)$. Incidentally, $\frac{\partial V}{\partial S}=0$ determines the VEV of $S$ as $\VEV{S}\sim \lambda^{s+2}/s^2$, which is quite small.

Finally, we have arrived at the following scenario;
\begin{eqnarray}
&m_{1 / 2}  \sim m_{3 / 2} \sim O(1 \mathrm{TeV}), \\
&m_{0}^2  \sim \psi g_A D_A \sim\left(10 m_{\text {SUSY }}\right)^2 \sim O\left((1000 \mathrm{TeV})^2\right)  ,\\
&m_{h}^2  \sim h g_A D_A \sim O\left((1000 \mathrm{TeV})^2\right)  ,
\end{eqnarray}
where $m_{0}^2$ and $m_h^2$ are mass square of sfermion with charge $\psi$ and the SUSY breaking Higgs mass square, respectively. 
We have assumed $m_{1/2}\sim1\rm{TeV}$ based on the experimental bounds, 
and thus the sfermion masses and the SUSY breaking Higgs mass become $O(1000)$ TeV. 
Since the sfermion masses are large, the SUSY contributions to FCNC processes and CP violating processes are strongly suppressed although the electroweak scale becomes unstable.

\subsection{Higgsino mass}
In a $D$-term dominant scenario, it is predicted that the Higgsino becomes much heavier than the weak scale, and therefore, it cannot be the dark matter. 
This is because in the natural GUT, the SUSY Higgs mass $\mu$, which is nothing but the Higgsino mass, must be taken to be large to cancel the large negative $D$-term contribution to $m_h^2$. 
Note that the Higgs $U(1)_A$ charge $h$ must not be zero. Actually, we have two reasons for negative $h$ in the natural GUT.
One of them is that the tree Higgs mass term 
$\lambda^{2h}H^2$ must be forbidden to solve the DT splitting problem. When $h<0$, it is forbidden by the SUSY zero mechanism. 
The other is that $h\sim -3$
\footnote{Although it differs from the requirements of GCU conditions, GCU can be achieved through fine-tuning of $O(1)$ coefficients.} 
is required to obtain the correct 
neutrino mass without any tuning of $O(1)$ coefficients \cite{Maekawa:2023sme}. 


If the $F$-term contribution becomes comparable with the $D$-term contribution, the Higgsino can be the dark matter when the $F$ and $D$ terms contributions to $m_h^2$ are cancelled out. 
Is it possible that the $D$-term contribution becomes comparable with the $F$-term contribution? That is our main question to be answered in this paper. 

\section{$F$-term or $D$-term?}
In this section, we discuss the possibility of avoiding the dominance of $D$-term contributions to the sfermion masses. Since the $D$-term dominance is a result in the simplest model, in this section we consider  more complicated models, in which the number of fields increases,   to discuss the possibility that the $D$-term dominance can be avoided.
Under the meta-stable vacuum, where the positively charged fields have quite small VEVs, we have no room to reduce the $D$-term contribution in the simplest model, which is determined by Eq. (\ref{Dfix}). However, 
if the number of positively charged fields increases, the number of non-negligible terms in the stationary condition as in Eq. (\ref{Dfix}) increases, and it may be possible to reduce the $D$-term contribution by a cancelation  between 
the non-negligible terms.  
Our goal is to clarify whether the Higgsino can be the dark matter or not in the natural GUT with spontaneous SUSY breaking. 
Therefore, we are interested in  the possiblity to make the $F$-term contribution and the $D$-term contribution comparable. 
To achieve this, $D_A$ must be reduced by two orders of magnitude.
Throughout this section, we do not assume $U(1)_R$ symmetry. However, we assume that the VEVs of positively charged fields are small and often neglect them. Whether this approximation is valid will be discussed at the end of this section.

\subsection{Demonstration with two positively charged fields}
We consider a simple model which has two positively charged fields and two negatively charged fields, as shown in the table \ref{simple}.
We take $s>t$ and $z<-1=\theta$ without loss of generality.

\begin{table}[t]
  \begin{center}
    \begin{tabular}{|c|c|}
      \hline
       negatively charged fields & positively charged fields  \\ \hline \hline
       $\Theta(\theta=-1)$ & $S(s>0)$  \\ \hline
       $Z(z<-1)$ & $T(t>0)$  \\ \hline
    \end{tabular}
    \caption{Field content of simple model.}
    \label{simple}
  \end{center}
\end{table}

Since the number of negatively charged fields and that of positively charged fields are equal, SUSY is spontaneously broken. In this model, the superpotential 
is given as
\begin{equation}
\label{W}
     W=a_{0} S\Theta^s +a_{1} SZ\Theta^{s+z}  +b_{0} T\Theta^t +b_{1} TZ\Theta^{t+z}, 
\end{equation}
where $a_{0},a_{1},b_{0}$ and $b_{1}$ are $O(1)$ coefficients.
Here, we neglected various terms for simplicity. For example, the terms which break $U(1)_R$ symmetry are neglected because in the meta-stable vacuum the positively charged fields have quite small VEVs. 
Furthermore, we neglect the terms which include more than one $Z$ fields for simplicity since these terms are not expected to play an important role in our argument 
because of $\lambda^{z}\langle Z\rangle=O(1)$.
The potential is given as
\begin{equation}
V=\left| F_S\right|^2+\left| F_T\right|^2+\left|F_Z \right|^2+\left|F_{\Theta}\right|^2+\frac{1}{2} D_A^2,
\end{equation}
where
\begin{eqnarray}
&& F_S^*=-a_0 \Theta^s-a_1 Z \Theta^{s+z} ,\\
&& F_T^*=-b_0 \Theta^t-b_1 Z \Theta^{t+z} ,\\
&& F_Z^*=-a_1 Z \Theta^{s+z}-b_1 T \Theta^{t+z} ,\\
&& F_\Theta^*=-s a_0 S \Theta^{s-1}-(s+z) a_1 S Z \Theta^{s+z-1}-t b_0 T \Theta^{t-1}-(t+z) b_1 T Z \Theta^{t+z-1} ,\\
&& D_A=g_A\left(\xi^2+s|S|^2+t|T|^2+z|Z|^2-|\Theta|^2\right).
\end{eqnarray}
Here, in the meta-stable vacuum,  SUSY breaking is caused by $F_S\sim \lambda^s$ because
$\lambda^s$ is smaller than $\lambda^t$.
We can understand the meta-stable vacuum by solving the stationary conditions
$\frac{\partial V}{\partial S}=0$,
$\frac{\partial V}{\partial T}=0$,
$\frac{\partial V}{\partial Z}=0$,
and $\frac{\partial V}{\partial \Theta}=0$. Basically, the first two stationary conditions, which are the derivative of the potential $V$ by the positively charged fields, determine the VEVs of positively charged fields, which are quite small. The last two stationary conditions, which are the derivative of the potential by the negatively charged fields, determine the VEVs of negatively charged fields, which are not so small. Therefore, in order to estimate the $D$-term, it is sufficient to consider the last two stationary conditions. Moreover, in the meta-stable vacuum, $F_T\sim 0$ because $\lambda^s\ll \lambda^t$, and therefore, $F_T\sim 0$ fixes approximately the VEV of $Z$.
We estimate $D$-term by the following steps;
\begin{enumerate}
\item Approximate $\VEV{\Theta}$ is determined by $D_A=0$ as $\VEV{\Theta}\sim\xi$, because
$\VEV{\Theta}$ is the largest VEV.
\item Calculate approximate $\VEV{Z}\equiv \VEV{Z}_0$ by F-flatness condition $F_T=0$ as
$\VEV{Z}_0\sim \lambda^{-z}$.
  \item Since $\VEV{Z}$ should be shifted due to SUSY breaking as
  $\VEV{Z}=\VEV{Z}_0+\Delta$, the shift $\Delta\ll\VEV{Z}_0$ is determined from the stationary condition $\frac{\partial V}{\partial Z}=0$.
  \item Substitute $\VEV{Z}$ into $\frac{\partial V}{\partial \Theta}=0$ and determine the order of $D_A$, (and correct $\VEV{\Theta}$ is fixed.)
\end{enumerate}

Now, we perform actual calculations.
The approximate VEV of negatively charged field is determined as $\VEV{Z}_0=-\frac{b_0}{b_1}\Theta^{-z}$ by $F_T=0$. Therefore, we can write $\VEV{Z}, F_S$ and $F_T$ as
\begin{eqnarray}
\label{Z}
&&\VEV{Z}=\left(-\frac{b_0}{b_1}+\delta\right)\Theta^{-z},\\
\label{FT}
&&F_T^*=-b_1 \delta \Theta^t, \\
\label{FS}
&&F_S^*=\left(-a_0+\frac{a_1 b_0}{b_1}-a_1 \delta\right) \Theta^s,
\end{eqnarray}
here $\delta\equiv \Delta\lambda^z$, where $\Delta$ is defined as the shift of $\VEV{Z}$ due to SUSY breaking, and is expected to be smaller than $1$.
We consider the following two stationary conditions;
\begin{eqnarray}
\label{V/Z}
\frac{\partial V}{\partial Z}&=&  F_S \frac{\partial F_S^*}{\partial Z}+F_T \frac{\partial F_T^*}{\partial Z}+D_A \frac{\partial D_A}{\partial Z} \nonumber\\
&=& \left(a_{0} \Theta^s+a_{1} Z\Theta^{s+z} \right)^*a_{1} \Theta^{s+z} \quad+\left(b_{0} \Theta^t+b_{1} Z\Theta^{t+z} \right)^*b_{1} \Theta^{t+z} \quad+z g_A D_A Z \nonumber\\
&=& 0 .\\
\label{V/theta}
\frac{\partial V}{\partial \Theta}&=&  F_S \frac{\partial F_S^*}{\partial \Theta}+F_T \frac{\partial F_T^*}{\partial \Theta}+D_A \frac{\partial D_A}{\partial \Theta} \nonumber\\
&=& \left(a_{0} \Theta^s+a_{1} Z\Theta^{s+z} \right)^*\left(sa_{0}  \Theta^{s-1}+(s+z)a_{1}Z \Theta^{s+z-1} \right) \nonumber\\
&& \quad+\left(b_{0} \Theta^t+b_{1} Z\Theta^{t+z} \right)^* \left(tb_{0}  \Theta^{t-1}+(t+z)b_{1}Z \Theta^{t+z-1} \right) \nonumber\\
&& \quad-g_A D_A \Theta\nonumber\\
&=&0 .
\end{eqnarray}
Here, we neglect terms which include $F_Z$ and $F_\Theta$
because they contain very small VEVs of $S$ and $T$.
The justification of this neglects will be discussed in the section III C, where the VEVs of positively charged fields are determined.
Note that a cancellation between $F_S \frac{\partial F_S^*}{\partial \Theta}$ and $F_T \frac{\partial F_T^*}{\partial \Theta}$ by tuning the $O(1)$ coefficients
$a_0$, $a_1$, $b_0$, and $b_1$ might make $D_A$ smaller than expected because these terms give the same order of contribution to $D_A$, that will be shown below.
This cancellation cannot be expected in the simplest model discussed in the previous section.
The shift parameter $\delta$ is determined by solving the stationary condition (\ref{V/Z})
as
\begin{equation}
\delta=-\frac{a_0^*a_1-|a_1|^2b_0/b_1}{|b_1|^2}\Theta^{2s-2t},
\end{equation}
by straightforward calculations.
Here, we can neglect the last term in (\ref{V/Z}) when $z<-1$.
Note that the first two terms in Eq. (\ref{V/theta}) become the same order by substituting 
$\delta\sim \lambda^{2s-2t}\ll 1$, and therefore the cancellation can be expected. 
Finally, $D_A$ can be determined by solving equation (\ref{V/theta}) as
\begin{equation}
\label{D_A}
    D_A=\frac{s}{g_A}\left|a_0-\frac{a_1 b_0}{b_1}\right|^2 \Theta^{2s-2},
\end{equation}
where we use $\delta\ll 1$.
Actually, if a relation 
\begin{equation}
  a_0b_1\sim a_1b_0 
  \label{relation}
\end{equation}
is satisfied, $D_A$ can be smaller than the expected value $s\lambda^{2s-2}$!
Unfortunately, even with the tuning condition (\ref{relation}), $D_A\sim s|F_S|^2\lambda^{-2}$ is still satisfied because
this relation (\ref{relation}) gives also smaller $F_S$, that is seen in Eq. (\ref{FS}). 
This result in this simple model suggests that the $D$-term contribution to sfermion mass squares cannot be smaller than $F$-term contribution by tuning the $O(1)$ coefficients. And therefore, the Higgsino dark matter is unlikely to occur in the natural GUT with spontaneous SUSY breaking.

\subsection{General models}
From the discussion so far, we have found that in a simple model, the $D$-term is proportional to the square of the $F$-term. Here, we will discuss whether this behavior can be confirmed in a more generalized model and the conditions for such confirmation.
In this time, we consider a model which has numerous positively charged fields $S,T_1,T_2,\dots,T_n$ and negatively charged fields $\Theta,Z_1,Z_2,\dots,Z_n$.
As before, we assume $s>t_i$, $\theta>z_i\ (i=1,2,\dots,n)$,  and $F_S$ causes SUSY breaking.
To obtain $D$-term, it is sufficient to consider $n+1$ stationary condition;
\begin{eqnarray}
&&\frac{\partial V}{\partial Z_i}=F_S \frac{\partial F_S^*}{\partial Z_i}+\sum_k F_{T_k} \frac{\partial F_{T_k}^*}{\partial Z_i}=0.\\
&&\frac{\partial V}{\partial \Theta}=F_S \frac{\partial F_S^*}{\partial \Theta}+\sum_k F_{T_k} \frac{\partial F_{T_k}^*}{\partial \Theta}-g_AD_{A} \Theta=0.
\end{eqnarray}
To perform the calculations, let us represent these in matrix form. They are wtitten as
\begin{eqnarray}
\label{V/Zmatrix}
&&F_S\left(\begin{array}{c}
\frac{\partial F_S^*}{\partial Z_1} \\
\vdots \\
\frac{\partial F_S^*}{\partial Z_n}
\end{array}\right)+\left(\begin{array}{ccc}
\frac{\partial F_{T_1}^*}{\partial Z_1} & \cdots &\frac{\partial F_{T_n}^*}{\partial Z_1}\\
\vdots & &\vdots  \\
\frac{\partial F_{T_1}^*}{\partial Z_n} & \cdots & \frac{\partial F_{T_n}^*}{\partial Z_n}\\
\end{array}\right)\left(\begin{array}{c}
F_{T_1} \\
\vdots \\
F_{T_n}
\end{array}\right)=0,\\
\label{V/thetamatrix}
&&F_S \frac{\partial F_S^*}{\partial \Theta}+\left(\frac{\partial F_{T_1}^*}{\partial \Theta} \cdots \frac{\partial F_{T_n}^*}{\partial \Theta}\right)\left(\begin{array}{c}
F_{T_1} \\
\vdots \\
F_{T_n}
\end{array}\right)-g_AD_{A} \Theta=0,
\end{eqnarray}
respectively.
From equation (\ref{V/Zmatrix}), we can write $F$-terms as
\begin{equation}
\label{FTmatrix}
\left(\begin{array}{c}
F_{T_1} \\
\vdots \\
F_{T_n}
\end{array}\right)=-F_S\left(\begin{array}{ccc}
\frac{\partial F_{T_1}^*}{\partial Z_1} & \cdots &\frac{\partial F_{T_n}^*}{\partial Z_1}\\
\vdots & &\vdots  \\
\frac{\partial F_{T_1}^*}{\partial Z_n} & \cdots & \frac{\partial F_{T_n}^*}{\partial Z_n}\\
\end{array}\right)^{-1}\left(\begin{array}{c}
\frac{\partial F_S^*}{\partial Z_1} \\
\vdots \\
\frac{\partial F_S^*}{\partial Z_n}
\end{array}\right).
\end{equation}
Furthermore, by substituting equation (\ref{FTmatrix}) into equation (\ref{V/thetamatrix}), we obtain
\begin{equation}
\label{V/thetamatrix2}
F_S\left[\frac{\partial F_S^*}{\partial \Theta}-\left(\frac{\partial F_{T_1}^*}{\partial \Theta} \cdots \frac{\partial F_{T_n}^*}{\partial \Theta}\right)\left(\begin{array}{ccc}
\frac{\partial F_{T_1}^*}{\partial Z_1} & \cdots & \frac{\partial F_{T_n}^*}{\partial Z_1} \\
\vdots & &\vdots\\
\frac{\partial F_{T_1}^*}{\partial Z_n} & \cdots & \frac{\partial F_{T_n}^*}{\partial Z_n}
\end{array}\right)^{-1}\left(\begin{array}{c}
\frac{\partial F_S^*}{\partial Z_1} \\
\vdots \\
\frac{\partial F_S^*}{\partial Z_n}
\end{array}\right)\right]-g_AD_A \Theta
=0.
\end{equation}
We consider a chain rule,  $\frac{d F_{T_i^*}}{d \Theta}=\frac{\partial F_{T_i}}{\partial \Theta}+\sum_k \frac{\partial F_{T_i}}{\partial Z_k} \frac{d Z_k}{d \Theta}$, where we take all VEVs of positively charged fields to be vanishing because they are expected to be very small. 
Since basically $F_{T_i}^*\sim 0$ are satisfied, $dF_{T_i}/d\Theta\sim 0$ are also satisfied. Note that this is not the case for $\partial F_{T_i}/\partial\Theta$, which can be large.
Thus, the above equation obtained by the chain rule becomes
\begin{equation}
\left(\frac{\partial F_{T_1}^*}{\partial \Theta} \cdots \frac{\partial F_{T_n}^*}{\partial \Theta}\right)=-\left(\frac{d Z_1}{d \Theta} \cdots \frac{d Z_n}{d \Theta}\right)\left(\begin{array}{ccc}
\frac{\partial F_{T_1}^*}{\partial Z_1} & \cdots & \frac{\partial F_{T_n}^*}{\partial Z_1} \\
\vdots & & \\
\frac{\partial F_{T_1}^*}{\partial Z_n} & \cdots & \frac{\partial F_{T_n}^*}{\partial Z_n}
\end{array}\right).
\end{equation}
Therefore, equation (\ref{V/thetamatrix2}) is rewritten as 
\begin{eqnarray}
0&=&F_S\left[\frac{\partial F_S^*}{\partial \Theta}+\left(\frac{d Z_1}{d \Theta} \cdots \frac{dZ_n}{d \Theta}\right)\left(\begin{array}{c}
\frac{\partial F_S^*}{\partial Z_1} \\
\vdots \\
\frac{\partial F_S^*}{\partial Z_n}
\end{array}\right)\right]-g_AD_A \Theta\nonumber\\
&=&F_S\frac{d F_S^*}{d \Theta}-g_AD_A\Theta,
\end{eqnarray}
and from this, it immediately follows that
\begin{equation}
D_A=\frac{1}{g_A}F_S\frac{d F_S^*}{d \Theta}\Theta^{-1},
\end{equation}
where $F_S^*=\alpha\Theta^s$ with a coefficient $\alpha$.
For any coefficient $\alpha$,  the above relation shows that $g_AD_A\sim s |F_S|^2\lambda^{-2}$. Therefore, we conclude that $D$-term dominance is a quite general result in the natural GUT with spontaneous SUSY breaking. 

Here, we have assumed that the VEVs of S and T are small and have omitted in the $F_{Z_i}$ terms and $F_\Theta$ term in the stationary conditions.
We study whether this assumption is correct or not in the next subsection.

\subsection{VEVs of positively charged fields}
In the above discussion, we assumed that the VEVs of $S$ and $T$, induced by SUSY breaking, are extremely small.
In this section, we use the model introduced in the section III A to demonstrate that such an assumption is valid.

First, we calculate the VEVs of positively charged fields. 
To do so, the second-order terms of positively charged fields are required. We set a superpotential
\begin{equation}
\begin{aligned}
W=S \Theta^s & + S Z \Theta^{s+z}+ S Z^2 \Theta^{s+2 z} \\
& + T \Theta^t+ T Z \Theta^{t+z}+ T Z^2 \theta^{t+2 z} \\
& + S T \Theta^{s+t}+ S^2 \Theta^{2 s}+ T^2 \Theta^{2 t}.
\end{aligned}
\end{equation}
In this subsection, we omit the $O(1)$ coefficients.
Higher-order terms of $S,T$ and $Z$ were omitted as their contributions do not change the order of the VEVs. 
$F$-terms are written as
\begin{eqnarray}
&&F_S^*=- \Theta^s- Z \Theta^{s+z}- Z^2 \Theta^{s+2 z}- T \Theta^{s+t}-2  S \Theta^{2 s},\\
&&F_T^*=- \Theta^t- Z \Theta^{t+z}- Z^2 \Theta^{t+2 z}- S \Theta^{s+t}-2  T \Theta^{2 t},\\
&&F_Z^*=- S \Theta^{s+z}-2  S Z \Theta^{s+2 z}- T \Theta^{t+z}+2  T Z \Theta^{t+2 z},\\
&&\begin{aligned}
F_\Theta^*=-s  S \Theta^{s-1} & -(s+z)  S Z \Theta^{s+z-1}-(s+2 z)  S Z^2 \Theta^{s+2 z-1} \\
& -t  T \Theta^{t-1}-(t+z)  T Z \Theta^{t+z-1}-(t+2 z)  T Z^2 \Theta^{t+2 t-1} \\
& -(s+t)  S T \Theta^{s+t-1}-2 s  S^2 \Theta^{2 s-1}+2 t  T^2 \Theta^{2 t-1}.
\end{aligned}
\end{eqnarray}
The possible largest-order terms of each are as follows;
\begin{equation}
F_S:\Theta^s,\quad F_T:\delta\Theta^t\sim\Theta^{2s-t},\quad F_Z:S\Theta^{s+z}+T\Theta^{t+z},\quad F_\Theta:S\Theta^{s-1}+T\Theta^{t-1}.
\end{equation}
The VEVs of positively charged fields $S$ and $T$ are determined by the stationary conditions 
\begin{eqnarray}
\frac{\partial V}{\partial S}&=&  F_S \frac{\partial F_S^*}{\partial S}+F_T \frac{\partial F_T^*}{\partial S}+F_Z \frac{\partial F_Z^*}{\partial S}+F_\Theta \frac{\partial F_\Theta^*}{\partial S}+D_A \frac{\partial D_A}{\partial S} =0,\\
\frac{\partial V}{\partial T}&=&  F_S \frac{\partial F_S^*}{\partial T}+F_T \frac{\partial F_T^*}{\partial T}+F_Z \frac{\partial F_Z^*}{\partial T}+F_\Theta \frac{\partial F_\Theta^*}{\partial T}+D_A \frac{\partial D_A}{\partial T} =0.
\end{eqnarray}
We can neglect the forth and fifth terms in the above equations because these terms give smaller contribution than the third term, 
that can be obtained from (\ref{D_A}).
Therefore, the two stationary conditions can be rewritten as the same relation, 
\begin{equation}
\Theta^{3s}+S\Theta^{2s+2z}+T\Theta^{s+t+2z}=0,
\end{equation}
excluding the $O(1)$ coefficients.
Finally, we can estimate the VEVs of positively charged fields as
\begin{eqnarray}
&& S\sim\Theta^{s-2z},\label{VEVS}\\
&& T\sim\Theta^{2s-t-2z}.\label{VEVT}
\end{eqnarray}

Second, we justify to neglect 
$F_Z$ and $F_\Theta$ in eqs. (\ref{V/Z}) and (\ref{V/theta}). 
The stationary condition for $Z$ which determines the shift of $\VEV{Z}$ is
\begin{eqnarray}
\frac{\partial V}{\partial Z}&=&  F_S \frac{\partial F_S^*}{\partial Z}+F_T \frac{\partial F_T^*}{\partial Z}+F_Z \frac{\partial F_Z^*}{\partial Z}+F_\Theta \frac{\partial F_\Theta^*}{\partial Z}+D_A \frac{\partial D_A}{\partial Z} \nonumber\\
&=&\Theta^{2s+z}+\Theta^{2s+z}+\Theta^{4s-z}+\Theta^{4s-3z-2}+\Theta^{2s-z-2}.
\end{eqnarray}
It is obvious that 
the first two terms dominate. This justifies the assumption made in the section III A and B. 
Furthermore, similar assumption for the stationary condition for $\Theta$ which determines $D_A$ is also justified in the same way. 


Finally, we can conclude that the $D$-term dominance is a fairly common result in the natural GUT with spontaneous SUSY breaking.

\section{Discussions and summary}

The natural GUTs solve the various problems in SUSY GUT by assuming that all terms allowed by $SO(10)\times U(1)_A$ symmetry are introduced with $O(1)$ coefficients.
The natural GUT also can naturally introduce the mechanism for spontaneous SUSY breaking. Therefore, the natural GUT with spontaneous SUSY breaking is one of the most attractive scenarios in the natural GUT. Unfortunately, this scenario predicts high scale SUSY, which causes the need for fine-tuning to realize the electroweak scale.
This is mainly because the cutoff scale becomes smaller than the Planck scale and partly because the $D$-term contribution dominates in the scenario.
Moreover, the $D$-term domination makes it impossible that the Higgsino is the lightest SUSY particle (LSP), and therefore, the Higgsino cannot be the dark matter. 

In this paper, we discussed whether it is possible to construct models in which the $D$-term contribution becomes smaller than the $F$-term contribution through the fine-tuning of $O(1)$ coefficients without changing the $F$-term.
As a result, it was shown quite generally that the $D$-term becomes larger than the square of the $F$-terms, making it difficult to fine-tune in a way that suppresses only the $D$-term.

In our calculation, we do not take into account the SUGRA effects, which may change our result.
By introducing the SUGRA effect, the $U(1)_R$ symmetry breaking is larger, and as a result, the VEVs of $S$ and $T$ become larger than those presented in this paper. 
This study may be important, but we think it is beyond the scope of this paper.

The dominance of the $D$-term suggests that the Higgsino is not a candidate for the dark matter if the $D$-term contribution to the Higgs mass square is negative, that is required to obtain correct values of neutrino masses and to solve the doublet-triplet splitting problem. However, it is interesting that the mass spectrum of sfermions is fixed solely by $D$-term contributions as in Ref. \cite{Maekawa:2019vzk}. We can test the various natural GUTs by measuring mass spectrum of sfermions.

\section{Acknowledgement}
This
work is supported in part by the Grant-in-Aid for Scientific Research from the Ministry of
Education, Culture, Sports, Science and Technology in Japan No. 19K03823(N.M.).

\end{document}